\documentclass{elsart}
\usepackage{epsfig,graphicx, amssymb}
\begin{document}
\runauthor{Litak et al.}

\begin{frontmatter}
\title{
Chaotic Vibration of a Quarter-Car Model Excited by the Road Surface Profile
}
\author[Lublin]{Grzegorz Litak\thanksref{E-mail},}
\author[Lublin]{Marek Borowiec,}
\author[Bristol]{Michael I. Friswell}
\author[Lublin]{Kazimierz Szabelski}

\address[Lublin]{Department of Applied Mechanics, Technical University of
Lublin,
Nadbystrzycka 36, PL-20-618 Lublin, Poland}
\address[Bristol]{
Department of Aerospace Engineering,
University of Bristol,
Queens Building,  Bristol BS8 1TR, United Kingdom
}

\begin{abstract}
  The Melnikov criterion is used to examine a global homoclinic 
bifurcation and transition to chaos in the case of a quarter car model 
excited kinematically by the road surface profile. By analyzing
the potential an analytic 
expression is found for the homoclinic orbit. By introducing an harmonic excitation 
term and damping as perturbations, 
the critical Melnikov amplitude of the road surface
profile is found, above which the system can vibrate chaotically. \\
 
\end{abstract}

\thanks[E-mail]{Fax: +48-815250808; E-mail:
g.litak@pollub.pl (G. Litak)}

\end{frontmatter}

\section{Introduction}

The problem of rough surface road profiles and its influence on vehicle unwanted vibrations
due to kinematic
excitations is still a subject of research among automotive manufacturers and research
groups, whose objective is
to minimize their effects on the driver and passengers
\cite{Verros2000,Gobbi2001,vonWagner2004,Turkay2005,Verros2005,Li2004}. Past
studies focused on the dynamics of a passive car suspension, the nonlinear 
characteristics
of tyres and the effect of shimming in vehicle wheels 
\cite{Szabelski1985,Szabelski1991,Mitschke1990}.
Recently many new applications of active and semi active control procedures and special 
devices to minimize  vehicle 
vibrations have been
developed \cite{Guo2004,Liu2005,Lauwerys2005}. Consequently old mechanical 
quarter 
car  models \cite{Szabelski1991,Mitschke1990}
have been re-examined in the
context of active damper applications. 
Dampers based on magnetorheological fluid with 
typical hysteretic characteristics have significant promise for 
effective vibration damping in many
applications \cite{Choi2001,Lai2002,Du2005}. 
New ideas in vehicle vibration damping, such as 'Sky hook' control
 \cite{Karnopp1974} or H$_{\infty}$ control 
\cite{Du2005,Soravia1996} have already been implemented and tested in 
several car and motorcycle applications. Efforts have focused on
studies of the excitation of the automobile by a road surface profile
with harmful noise components  \cite{vonWagner2004,Turkay2005}.   
However, due to various nonlinearities in the vehicle dynamics, chaotic behaviour may 
produce noise like responses \cite{Li2004,Zhu2004,Zhu2006}.  

In the present paper the model of Li \textit{et al.} \cite{Li2004} is used, with the 
addition of a gravitational term that changes the equilibrium point and therefore the external potential.
This paper uses the Melnikov theory 
\cite{Melnikov1963,Guckenheimer1983,Tyrkiel2005} 
to estimate the critical amplitude of the road surface
profile above which the system can vibrate chaoticaly.
 
The gravitational term changes the topology of the heteroclinic orbit (in the case of 
a symmetric reversed 'Mexican 
hat' potential) into a homoclinic one (with a broken symmetry potential).
Systems with Duffing characteristics having non-symmetric potentials 
and a linear repulsive force term
are a wide class of mechanical systems and have been the
subject of previous investigations in the context of the appearance of chaotic solutions 
\cite{Cicogna1987,Brunsden1989,Lenci2004,Litak2005a}.  This paper
uses a similar approach for a quarter car model with a magnetorheological damper 
\cite{Choi2001,Lai2002,Du2005}.

\section{The quarter-car model}

\begin{figure}[htb]
 \centerline{
 \epsfig{file=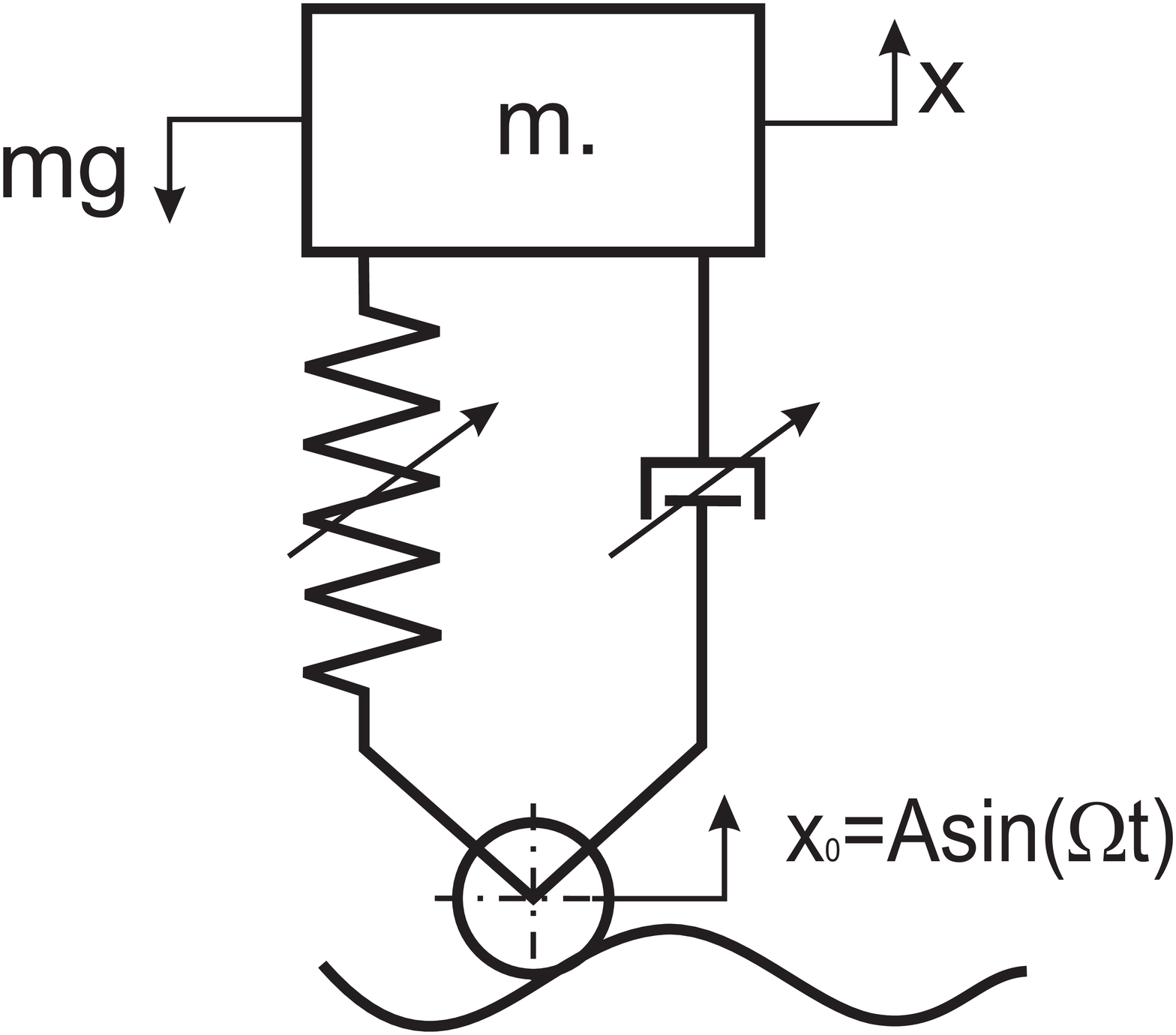,width=6.5cm,angle=0}}
 \caption{
   \label{fig1}
   The quarter-car model subjected to kinematic excitation with nonlinear
   damping and stiffness.
 }
\end{figure}

The equation of motion of a single degree of freedom quarter-car 
model (Fig. \ref{fig1}) is \cite{Li2004}
\begin{equation}
 \label{eq1}
 m \frac{{\rm d}^2 x}{{\rm d} t^2} +k_1(x-x_0)+mg+
 F_h\left( \frac{{\rm d}}{{\rm d} t}(x-x_0),x-x_0\right) = 0,
\end{equation}
where $F_h$ is an additional 
nonlinear hysteretic  suspension damping and 
stiffness force dependent on relative displacement and velocity, given by
\begin{eqnarray}
 && F_h\left(\frac{{\rm d}}{{\rm d} t}(x-x_0),x-x_0\right) = k_2(x-x_0)^3+
 c_1 \frac{{\rm d}}{{\rm d} t} \left( x-x_0 \right) +
 c_2 \left( \frac{{\rm d}}{{\rm d} t} \left( x-x_0 \right) \right)^3 \nonumber
 \\ &&  \label{eq2}
\end{eqnarray}
and 
\begin{equation}
 \label{eq3}
 x_0=A \sin (\Omega t).
\end{equation}
Defining a new variable for a relative displacement as
\begin{equation}
 \label{eq4}
 y=x-x_0
\end{equation}
we get
\begin{equation}
\label{eq5}
 \frac{{\rm d}^2 y}{{\rm d} t^2} + \omega^2 y + B_1 y^3 +
 B_2 \frac{{\rm d} y}{{\rm d} t} + 
 B_3 \left( \frac{{\rm d} y}{{\rm d} t} \right)^3 =
 - g + A\Omega^2\sin(\Omega t),
\end{equation}
where $\omega^2=k_1/m$, $B_1=k_2/m$, $B_2=c_1/m$, $B_3=c_2/m$.

Following Li \textit{et al.} \cite{Li2004} the system parameters are defined as
\begin{eqnarray}
 \label{eq6}
 m=240~{\rm kg}, ~k_1=160 000~{\rm N/m}, ~k_2=-300 000~{\rm N/m}^3, \\
 c_1=-250~{\rm Ns/m}, ~c_2=25~{\rm Ns}^3{\rm /m}^3. \qquad \qquad \nonumber
\end{eqnarray}

The corresponding dimensionless equation of motion can be written for a scaled time 
variable $\tau=\omega t$ as:
\begin{equation}
 \label{eq7}
 \ddot y + y + ky^3 + \alpha \dot y + \beta \dot y^3 = - g' + A \Omega'^2 \sin (\Omega' \tau),
\end{equation}
where
$k=B_1/\omega^2=\frac{k_2}{k_1}$, $\alpha=B_2/\omega=\frac{c_1}{\sqrt{k_1 m}}$, 
$\beta=B_3~\omega=c_2\sqrt{\frac{k_1}{m^3}}$, $g'=\frac{g}{\omega^2}$, $\Omega'=\Omega/\omega$. 
The overdots in Eq.~(\ref{eq7}) denote the corresponding derivative with respect to $\tau$ 
(~${\bf \dot{~}} ~\equiv {\rm d}/{\rm d} \tau$).


Note that in our model, Eq. (\ref{eq5}), we use both a complicated non-symmetric potential
and also non-trivial damping of the Rayleigh type. Similar damping terms
have been used  before in the context of Melnikov theory  \cite{Litak1999,Trueba2000}.
Litak {\em et al.}  
\cite{Litak1999} considered the Froude pendulum, with polynomial damping to model 
a dry friction phenomenon. 
Trueba {\em et al.}  \cite{Trueba2000}
performed systematic studies for basic nonlinear oscillators including those with combined 
damping.   
Here the motivation in using a complicated damping term is different, and 
arises from the use of magnetorheological dampers in 
vehicle suspensions \cite{Choi2001,Lai2002,Du2005}. 
The signs of the $c_i$ coefficients (Eqs. \ref{eq1}-\ref{eq2}) have changed 
compared to reference 
\cite{Li2004}, in order to recover the usual Rayleigh term $c_2v^3+c_1v$, 
where $v={\rm d} y/{\rm d} t$ is the 
system velocity and $c_1 < 0$, $c_2 > 0$ (defined Eq. (\ref{eq6})).  
This term is able to drive the system into a stable limit cycle
solution, being dissipative for a large enough velocity $v$ $\left( v > 
\sqrt{c_1/c_2} \right)$
and pumping energy for a small velocity $\left( v < \sqrt{c_1/c_2} \right)$.

In Eq.~(\ref{eq7}), the nonlinear stiffness force has the potential
\begin{equation}
\label{eq8}
 V(y) = g'y + \frac{1}{2}y^2 + \frac{k}{4}y^4.
\end{equation}
Figure \ref{fig2} shows this potential, and highlights the characteristic fixed points. 
Note the non-symmetry is caused by the gravitational term $g'y$, and that $k<0$.

In Figs. \ref{fig3}a-b we show the results of calculations in the interesting region
of the main resonance for the system parameters given in Eq.~(\ref{eq6}) and a 
realistic amplitude of road profile excitation, namely $A=0.11$~m.  
In this case the vehicle vibration amplitude, $A_{OUT}$, plotted in Fig. \ref{fig3}a, 
has been  
determined numerically. For simplicity it has been  defined as:
\begin{equation}
\label{eq9}
 A_{OUT}=(y_{max}-y_{min})/2,
\end{equation}
where $y_{max}$ and $y_{min}$ are the maximum and minimum response of the vehicle model in 
the steady state. The resonance curve was calculated by tracking the solution for 
decreasing $\Omega'$, and indicates that the main resonance occurs at $\Omega' \approx 0.85$. The 
response curve is 
inclined to the left, as expected for a nonlinear system with softening stiffness 
characteristic.
Clearly, a jump between large and small vibration amplitudes exists at 
$\Omega' \approx 0.8$. Below this frequency we also observe a second, but much smaller, maximum of 
$A_{OUT}$ (at $\Omega' \approx 0.75$) which indicates that 
something interesting is occurring at this frequency. To examine this effect in more detail 
Fig. \ref{fig3}b shows a 
bifurcation diagram over the same range of excitation frequencies.
Interestingly $\Omega' \approx 0.75$ is a point of dramatic change in the system behaviour. 
For any frequency below 
this point we see a black bounded region while above it there are singular points.
One can easily see that the local change in $A_{OUT}$ is associated with a Hopf bifurcation. This 
transition is usually connected with a synchronization phenomenon between the system vibration 
frequency and the external excitation frequency. 
It is also connected with a slight change of the size of the attractor, reflected in the plot of $A_{OUT} 
(\Omega')$ (Fig. \ref{fig3}a).

To show the changes in the dynamics caused by this resonance and Hopf bifurcation, 
Figs. \ref{fig4}a-c show the phase portraits and
Poincare maps of the system for three chosen frequencies. 
One can easily identify Fig. \ref{fig4}a ($\Omega'= 0.6$) as a quasi-periodic 
solution with a limit cycle attractor. On the other hand the
solutions presented in Fig. \ref{fig4}b ($\Omega'= 0.8$) and Fig. \ref{fig4}c ($\Omega'= 1.1$)   
show synchronized motion represented by singular points. The range of the vibration 
amplitudes is the highest for the last 
case examined ($\Omega'= 1.1$), which is consistent with Fig. \ref{fig3}a.

Figure \ref{fig4}d shows the hysteresis of the function $F_h(\dot y, y)$, 
defined by Eq.~(\ref{eq2}) and obtained during the same simulation sweeps as the phase 
portraits.
The corresponding hysteresis loops differ in size. Starting with '1' plotted for $\Omega'= 0.6$
then increasing strongly in size ('2' plotted for $\Omega'= 0.8$) and finally decreasing
('3' plotted for $\Omega'= 1.1$). Note this sequence differs from the changes in the 
vibration 
amplitude, where $\Omega'= 1.1$ has the largest amplitude, $A_{OUT}$, and  
this arises because of the combined effect of displacement and velocity.

\section{Melnikov Analysis}

\begin{figure}[htb]
 \centerline{
 \epsfig{file=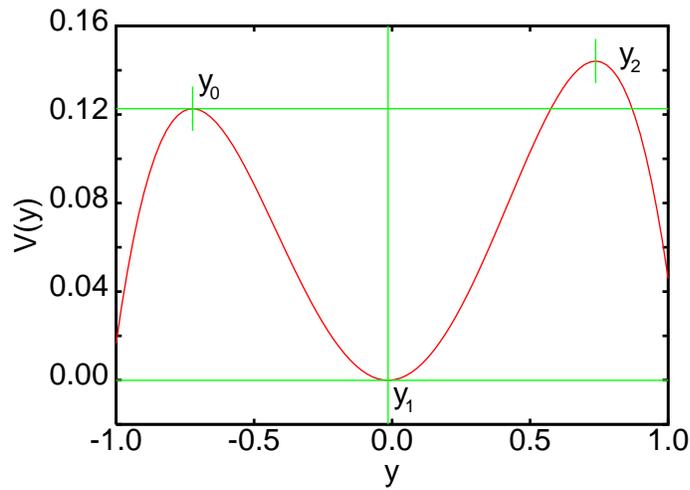,width=7.5cm,angle=-90}}
 \caption{ \label{fig2}
  External potential $V(y)$ (Eq. (\ref{eq8})) for given system parameters
  (Eq. (\ref{eq6})). $V(y)$ is scaled in Nm while $y$ is in m.
  The fixed points are $y_0=-0.7228$m, $y_1=-0.0147$m, $y_2=0.7375$m.
 }
\end{figure}

\begin{figure}[htb]
 \centerline{
 \epsfig{file=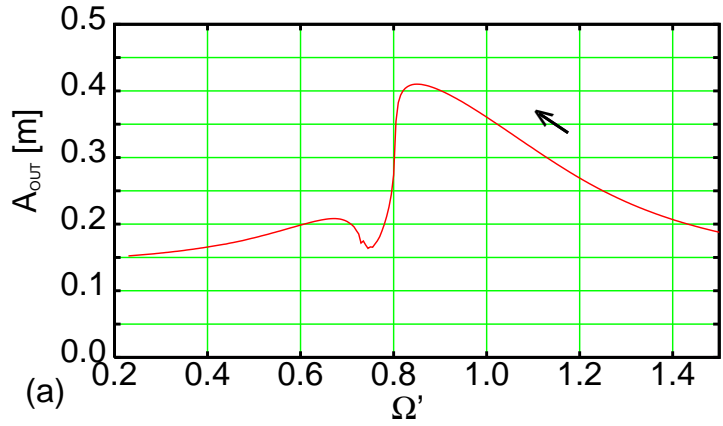,width=6.5cm,angle=-90}}
 \vspace{1cm} ~~\\
 \centerline{
 \epsfig{file=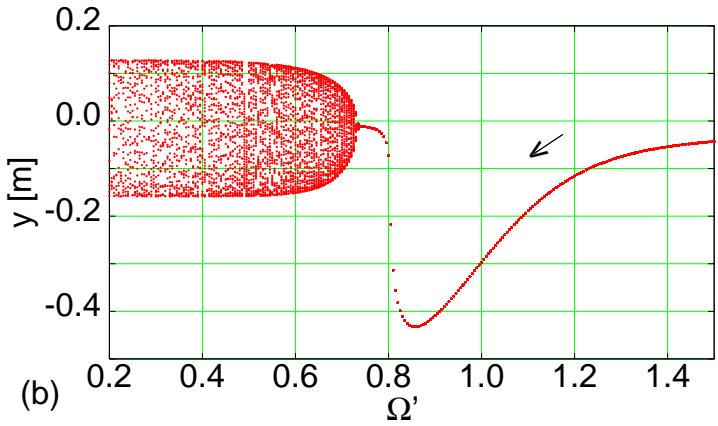,width=6.5cm,angle=-90}
 }
 \caption{ \label{fig3}
 Vibration amplitude $A_{OUT}=(y_{max}-y_{min})/2$ (Fig. \ref{fig3}a) and
 bifurcation diagram (Fig. \ref{fig3}b). The amplitude of a road profile
has been taken as
 $A=0.11$m. The arrows indicates $\Omega'$ reduces in the simulations.
For each new smaller $\Omega'$ the initial conditions $[y_{in},v_{in}]$
were taken as the
final position and velocity for the previous $\Omega'$.
 }
\end{figure}

\begin{figure}[htb]
 \centerline{
 \epsfig{file=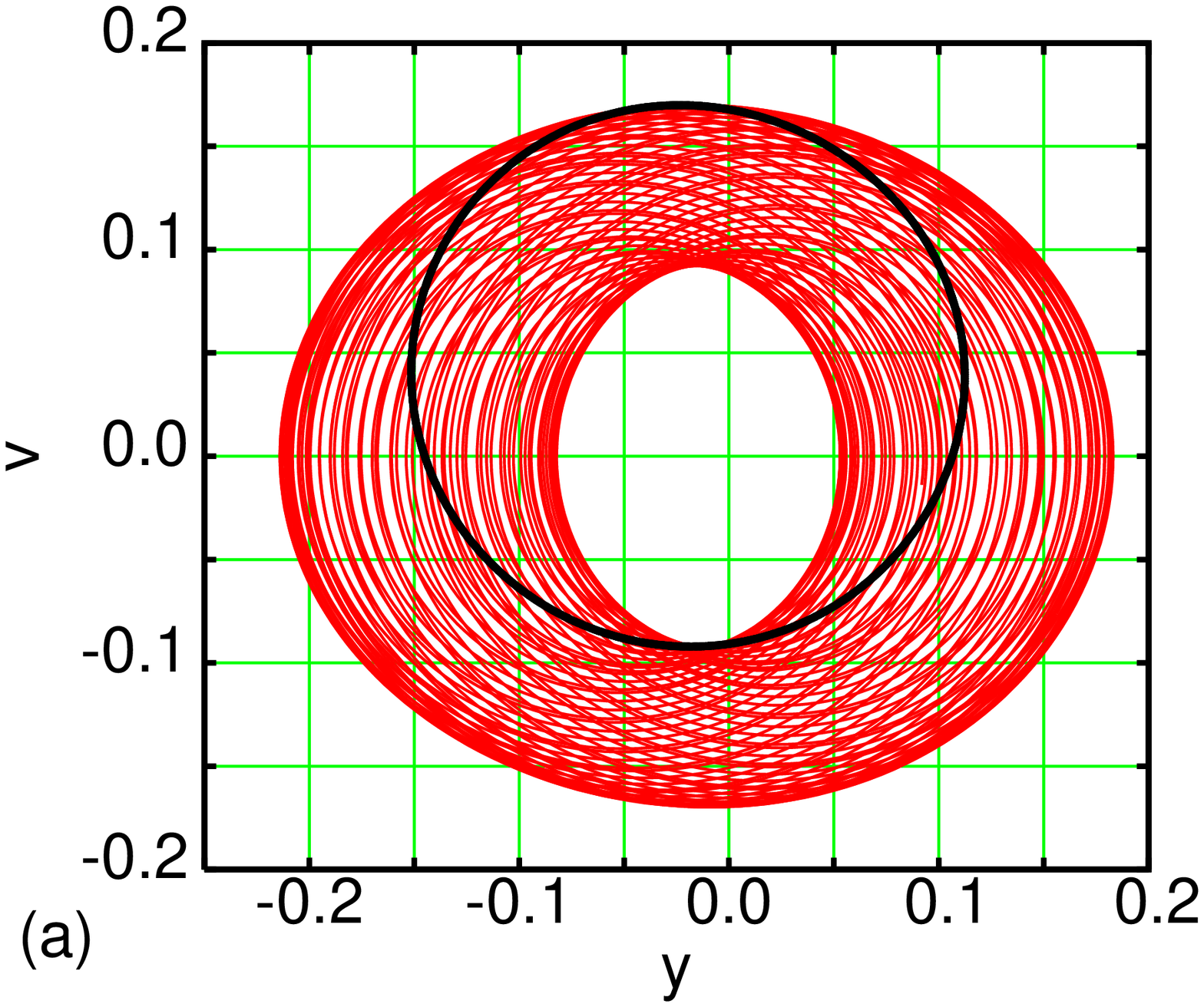,width=6.5cm,angle=-90}
 \epsfig{file=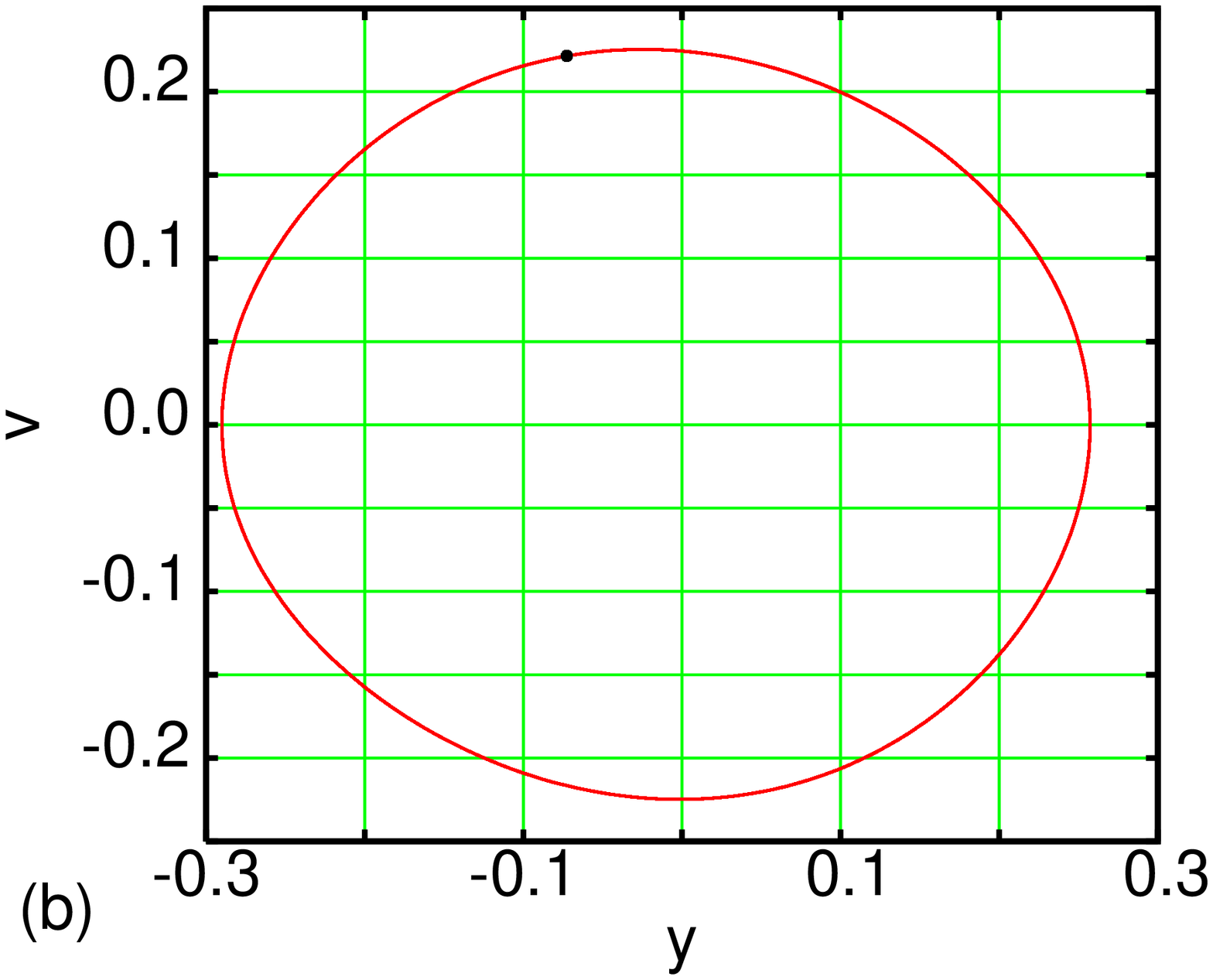,width=6.5cm,angle=-90}
 }
 \centerline{
 \epsfig{file=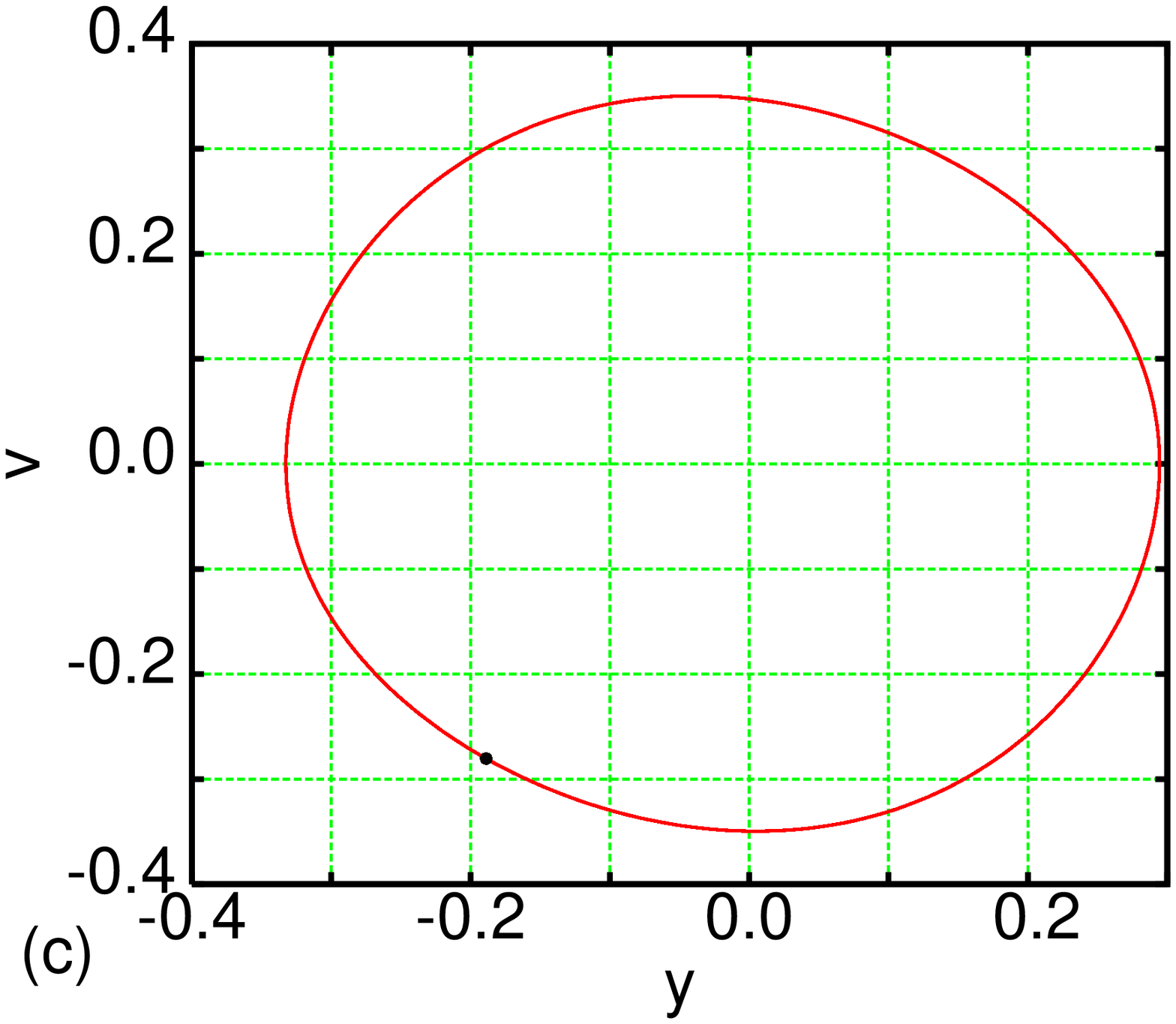,width=6.5cm,angle=-90}
 \epsfig{file=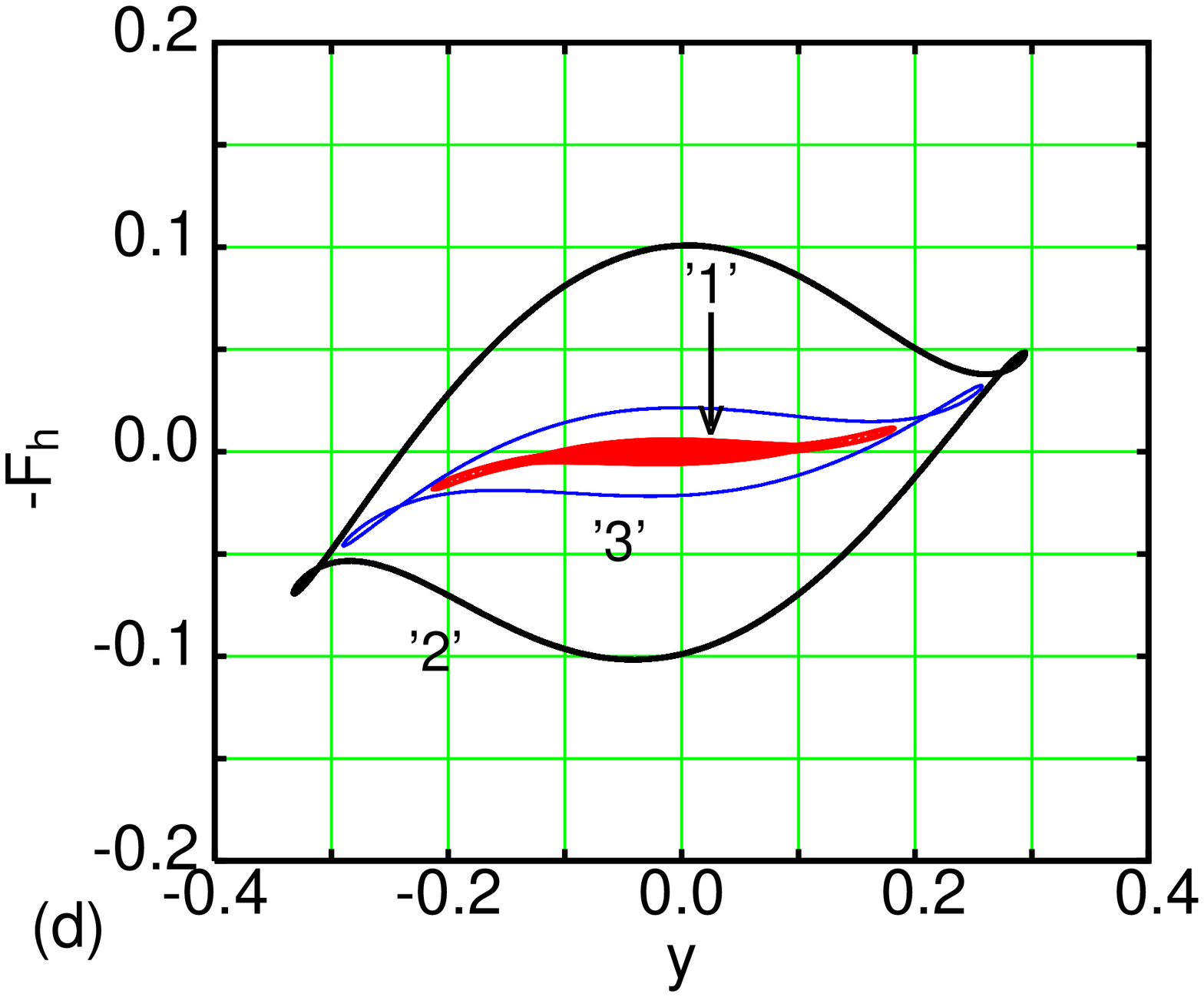,width=6.5cm,angle=-90}
 }
 \caption{ \label{fig4}
  Phase diagrams (velocity $v=\dot y$ versus displacement $y$ plotted by lines)
  and corresponding Poincare sections (plotted by  points) for $A=0.11$m and
  different $\Omega'$: $\Omega'=0.6$ (Fig. \ref{fig4}a),
  $\Omega'=0.8$ (Fig. \ref{fig4}b) and $\Omega'=1.1$ (\ref{fig4}c). The
  corresponding hysteresis curves are shown in Fig. \ref{fig4}d, where
   '1', '2' and '3' represent
  $\Omega'=0.6$, 0.8 and 1.1, respectively. $v \omega$ is scaled in m/s, $y$ in m,
  while the renormalized $F_h$ is presented in dimensionless units
  $F_h=\alpha \dot y+\beta \dot y^3+ky^3$ (see Eq. (\ref{eq7})).
 }
\end{figure}

\begin{figure}[htb]
 \centerline{
 \epsfig{file=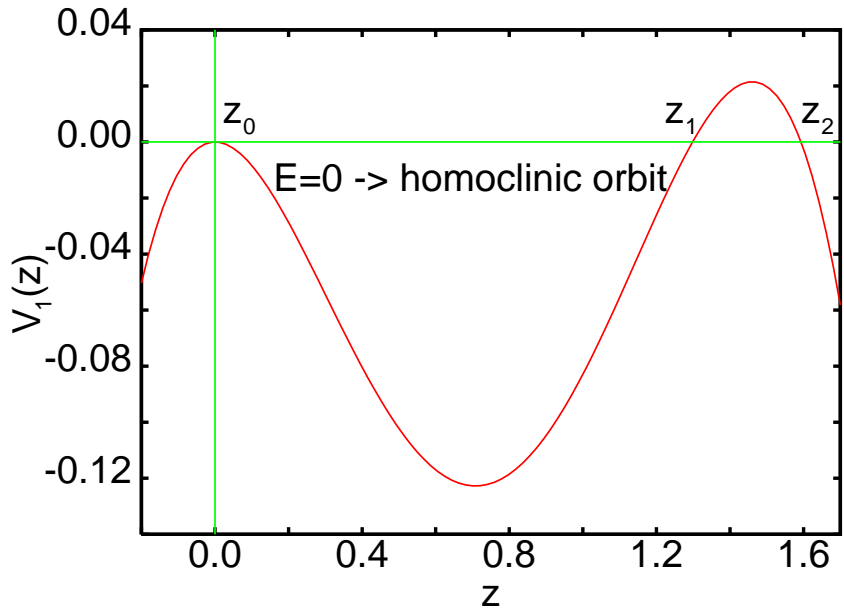,width=9.5cm,angle=-90}}
 \caption{ \label{fig5}
  Renormalized external potential $V_1(z)= \frac{k}{4}z^2(z-z_1)(z-z_2)$
  for given system parameters (in our case: $k= -1.875$ N/m$^3$,
  $z_1 \approx 1.298$m and $z_2 \approx 1.593$m).
  $z$ is expressed in m while $V(z)$ is in Nm.
  }
\end{figure}

Melnikov analysis starts with the renormalisation of the potential (Eq. (\ref{eq8}), 
Fig. \ref{fig2}) \cite{Litak2005a}. If we let $y = z + y_0$, where 
$y_0$ is the fixed point given in Fig.~\ref{fig2}, and $V_1(z) = V(y) - V(y_0)$, then,  
\begin{equation}
 \label{eq10}
 V_1(z)=\frac{k}{4} z^2(z-z_1)(z-z_2), 
\end{equation} 
where $z_1 = 1.298$ and $z_2=1.593$. Fig. \ref{fig5} shows this normalized potential. 
Notice that the left peak (the saddle point) of the potential $V_1(z)$ occurs at 
$z=0 < z_1 < z_2$ and that $V_1(0)=0$.

Looking for a homoclinic orbit we introduce a small parameter
$\epsilon$ (formally $\epsilon \tilde \alpha = \alpha$, $\epsilon \tilde \beta = \beta$ 
 and $\epsilon \tilde A 
=A$). The 
equation of motion then has the following form,
\begin{equation}
\ddot{z} +\epsilon  \tilde \alpha \dot{z} +\epsilon \tilde \beta \dot{z}^3
+k\left( z^3- \frac{3}{4}(z_1+z_2)z^2 + \frac{1}{2}z_1z_2z \right) = \epsilon 
\tilde A  \Omega'^2
\sin{(\Omega' \tau)}.  \label{eq11}
\end{equation}

Rewriting this second order differential equation 
as two first order differential equations yields,
\begin{eqnarray}
& & \dot{z}= v,  \label{eq12}
\\
& & \dot{v}= -kz^3+ \frac{3k}{4}(z_1+z_2)z^2
  -\frac{k}{2}z_1z_2z +\epsilon (-\tilde \alpha v -\tilde \beta v^3
  + \tilde A \Omega'^2
\sin{(\Omega' \tau)}). \nonumber
\end{eqnarray}

Note that the unperturbed equations (for $\epsilon=0$)
can be obtained from the gradients of the Hamiltonian $H^0(z,v)$,
\begin{equation}
 \dot{z}= \frac{\partial H^0}{\partial v}, ~~~~~\dot{v}= -\frac{\partial H^0}{\partial z},
 \label{eq13}
\end{equation}  
where $H^0$ is defined as
\begin{equation}
 \label{eq14}
H^0= \frac{v^2}{2} + \frac{k}{4}(z-z_1)(z-z_2)z^2.
\end{equation}

The homoclinic orbits are obtained from the unperturbed Hamiltonian, 
Eq. (\ref{eq14}), as
\begin{equation}
\label{eq15}
\tau= \sqrt{\frac{2}{-k}}\int \frac{{\rm d} z }{z\sqrt{(z-z_1)(z-z_2)}}.
\end{equation}
which may be evaluated in the following form:
\begin{equation}
\label{eq16}
\tau-\tau_0= \sqrt{\frac{2}{-z_1z_2k}}
\ln{\left|\frac{2z_1z_2-(z_1+z_2)z+2\sqrt{z_1z_2(z-z_1)(z-z_2)}}{z}\right|},
\end{equation}
where $\tau_0$ is a time like constant of integration.

\begin{figure}[htb]
 \centerline{
 \epsfig{file=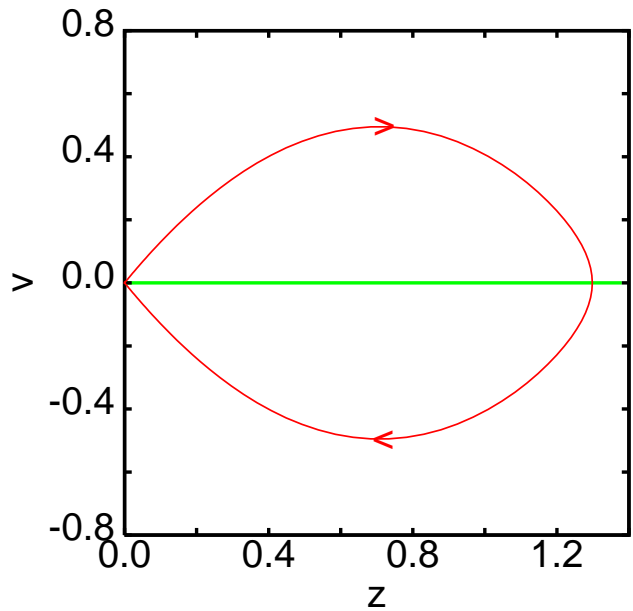,width=9.5cm,angle=-90}} 
 \caption{ \label{fig6}
A homoclinic orbit for the given potential (Eq. (\ref{eq9}) and  Fig. \ref{fig5}).
$z$ is expressed in m while $v \omega$ is given in m/s.
}  
\end{figure}

Thus, the single homoclinic orbit is given by the inverse of the above
expression and
the corresponding velocity ($z^*(\tau-\tau_0),v^*(\tau-\tau_0)$), as \cite{Litak2005a},
 \begin{eqnarray}
&& \hspace{-1cm} 
z^*=\frac{4z_1z_2 \exp\left((\tau-\tau_0)\sqrt{\frac{-kz_1z_2}{2}}
\right)}{-(z_1-z_2)^2 - \exp\left(2(\tau-\tau_0)\sqrt{\frac{-kz_1z_2}{2}}
\right) +2(z_1+z_2) \exp\left((\tau-\tau_0)\sqrt{\frac{-kz_1z_2}{2}}
\right)}   \nonumber
\\
&&
  \nonumber \\
&& \hspace{-2cm}
\label{eq17} \\
&&  
  \nonumber \\
&& \hspace{-1cm}  
v^*=\frac{-4z_1z_2
\sqrt{\frac{-kz_1z_2}{2}}\exp\left((\tau-\tau_0)\sqrt{\frac{-kz_1z_2}{2}}
\right)\left(
(z_1-z_2)^2-\exp\left(2(t-t_0)\sqrt{\frac{-kz_1z_2}{2}}
\right) \right)
}{\left(-(z_1-z_2)^2 - \exp\left(2(\tau-\tau_0)\sqrt{\frac{-kz_1z_2}{2}}
\right) +2(z_1+z_2) \exp\left((\tau-\tau_0)\sqrt{\frac{-kz_1z_2}{2}}
\right)\right)^2}.  \nonumber
\end{eqnarray}
Now suppose that
\begin{equation}
\label{eq18}
 \tau_0=\tau_{01}+\tau_{02}, \quad \textrm{where} 
 ~~\tau_{01}=-\ln\left(\frac{\sqrt{2} (z_2-z_1)}{\sqrt{-kz_1z_2}} \right).
\end{equation}
$\tau_{01}$ has been fixed to guarantee the  
proper parity (under the time transformation $\tau \rightarrow -\tau$), and hence
\begin{equation} \label{eq19} 
  z(-\tau)=z(\tau)
  \quad \textrm{and} \quad
  v(-\tau)=-v(\tau).
\end{equation}
$\tau_{02}$ is an arbitrary constant to be determined later in the minimization 
of the Melnikov integral 
${\rm M}(\tau_{02})$.  
The corresponding orbit 
($z^*(\tau-\tau_0)$,$v^*(\tau-\tau_0)$) is  
plotted on the phase plane in Fig. \ref{fig6}.

The distance between perturbed stable and unstable manifolds and their 
possible cross-sections may be examined by means of
the Melnikov integral ${\rm M}(\tau_{02})$, given by
\cite{Guckenheimer1983}
\begin{eqnarray}
\label{eq20}
{\rm M}(\tau_{02}) = \int_{- \infty}^{ + \infty}  {\rm {\bf h}}( 
z^*(\tau-\tau_{01}-\tau_{02}), 
v^*(\tau-\tau_{01}-\tau_{02})) \qquad \\  
 \wedge  {\rm {\bf g}} ( 
z^*(\tau-\tau_{01}-\tau_{02}),
v^*(\tau-\tau_{01}-\tau_{02})) {\rm d} \tau, \nonumber
\end{eqnarray}
where the wedge product for two dimensional vectors is defined 
as ${\bf h} \wedge {\bf g} = h_1g_2 - h_2g_1$.
The corresponding vector ${\bf h}$ is the gradient of 
unperturbed Hamiltonian (Eq. \ref{eq13}),
\begin{equation}
\label{eq21}
{\bf h} =  
\left[k(-z^{*3}+\frac{3}{4}(z_1+z_2)z^{*2}-\frac{1}{2}z_1z_2z^*), 
v^*\right],
\end{equation}
while the vector ${\bf g}$ consists of the perturbation terms to the same 
Hamiltonian (Eq. (\ref{eq10})),
\begin{equation}
\label{eq22}
{\bf g} =
\left[-\tilde \alpha v^* -\tilde \beta v^{*3}+
 \tilde
A \Omega'^2
\sin{(\Omega' \tau)}, 0 \right].
\end{equation}

Thus, shifting the time coordinate $\tau \rightarrow \tau+\tau_{02}$ under the integral (Eq. 
(\ref{eq20})), gives
\begin{eqnarray}
{\rm M}(\tau_{02}) &=& 
 \int_{-\infty}^{\infty} v^*(\tau-\tau_{01})\left((-\tilde \alpha v^*(\tau-\tau_{01})
-\tilde \beta v^{*3}(\tau-\tau_{01}) \right. \nonumber \\
 &+& \left. \tilde A \Omega'^2 \sin{(\Omega'( \tau+\tau_{02}))}\right) {\rm d} \tau. 
\label{eq23}
\end{eqnarray}

Finally, 
a sufficient condition for a global homoclinic transition
corresponding to a
horseshoe type of stable and unstable manifold
cross-section (for the excitation amplitude $A >
A_c$), can be written as:
\begin{equation}
\label{eq24}
{\displaystyle \bigvee_{\tau_{02}}}
~~~ {\rm M}(\tau_{02})=0 {\rm ~~ and ~~}
\frac{\partial {\rm M}(\tau_{02})}{\partial \tau_{02}} \neq 0.
\end{equation}

From  Eqs. (\ref{eq23}) and (\ref{eq24})
\begin{equation}
\label{eq25}
A_c= \frac{I_1}{\Omega'^2 I_2 (\Omega')},
\end{equation}
where
\begin{equation}
\label{eq26}
I_1 = \left| \int_{-\infty}^{\infty}
(\alpha (v^*(\tau))^2 +\beta (v^*(\tau))^4 ){\rm d} \tau \right|
\end{equation}
and 
\begin{eqnarray}
\label{eq27}
&& I_2 (\Omega')
= \sup_{\tau_{02} \in {\rm R} } 
\left| \int_{-\infty}^{\infty} v^*(\tau-\tau_{01}) \sin{(\Omega' 
(\tau+\tau_{02}))} 
{\rm 
d} 
\tau \right| \\ &&
=  \left|\int_{-\infty}^{\infty} v^*(\tau-\tau_{01}) \sin{(\Omega' \tau)} 
 {\rm d} \tau \right|, \nonumber 
\end{eqnarray}
where sup means supremum for various $\tau_{02}$, and is practically realized by
\begin{equation}
 \label{eq28}
 \cos{(\Omega' \tau_{02})}= \pm 1.
\end{equation}

Equation \ref{eq27} has been obtained
by using a trigonometric identity:
  $\sin(\psi_1+\psi_2) = \sin(\psi_1)\cos(\psi_2) + \cos(\psi_1)\sin(\psi_2)$
where $\psi_1=\Omega' \tau$ and $\psi_2=\Omega' \tau_{02}$.
We left only the term $\cos(\psi_2)$  (Eq. \ref{eq28}) because of the odd parity of 
the velocity function $v^*(\tau-\tau_{01})$ 
under
the integral. Of course to do such a simplification we needed defined
parity of $v^*$ (Eq. \ref{eq20}) 
and a proper choice of a constant $\tau_{01}$ (Eq. \ref{eq19}).

\begin{figure}[htb]
\centerline{
\epsfig{file=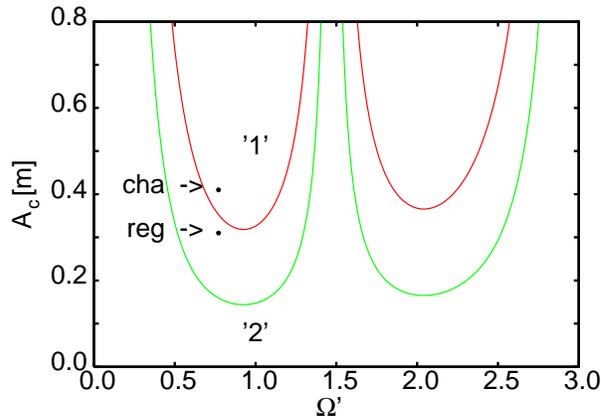,width=6.5cm,angle=-90}
}
\caption{ \label{fig7}
Critical amplitude $A_c$ versus vibration frequency for two different damping choice.
The damping coefficients $c_2=25$ Ns$^3$/m$^3$ (for other system parameters see Eq.
(\ref{eq6})) , and $c_2=12.5$
Ns$^3$/m$^3$, for '1' and '2' curves respectively.
The points 'reg' and 'cha' denote regular and
chaotic solutions (to be examined later in Figs. \ref{eq9}a and b).
}
\end{figure}

Note, the above integrals  may be evaluated analytically \cite{Schmidt1994} but here, for 
simplicity, they are
calculated numerically \cite{Cicogna1987,Litak2005a}.
Figure \ref{fig7} shows $A_c$ as a function of $\Omega'$ for $\alpha \cong -0.04$ 
and  $\beta\cong 2.69$ (see Eqs. 
(\ref{eq6}) and (\ref{eq7})) given by the curve labelled '1' and $\beta\cong 2.69/2$ given by the 
curve labelled '2'. 
One can see the characteristic double sack-like shape, similar to the structure 
observed by Lenci and Rega \cite{Lenci2004}.
This structure is governed by 
the oscillating term $\sin{(\Omega' \tau)}$ in the denominator of the integral $I_2 (\Omega')$ 
(Eq. (\ref{eq25})).

\section{Results of Simulations}

\begin{figure}[htb]
\centerline{
\epsfig{file=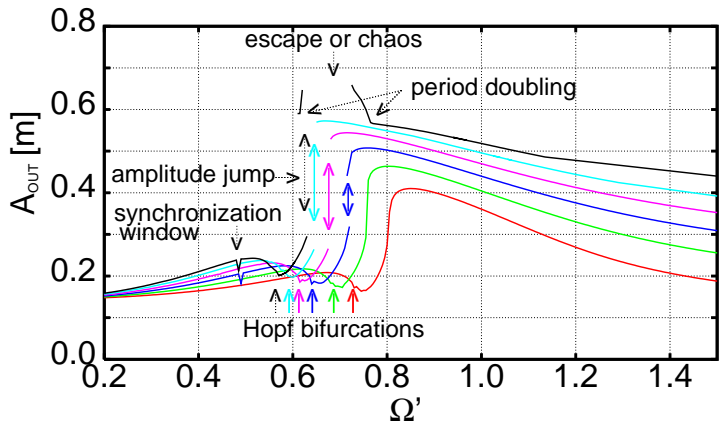,width=6.5cm,angle=-90}   
}

\caption{ \label{fig8}
Sequence of vehicle vibration amplitudes $A_{OUT}=(y_{max}-y_{min})/2$ versus frequency $\Omega'$
for road profile amplitudes of $A=0.11$, 0.16, 0.21, 0.26, 0.31 and 0.36m
(from the lower to upper curves, respectively).
}
 \end{figure}

\begin{figure}[htb]
\centerline{
\epsfig{file=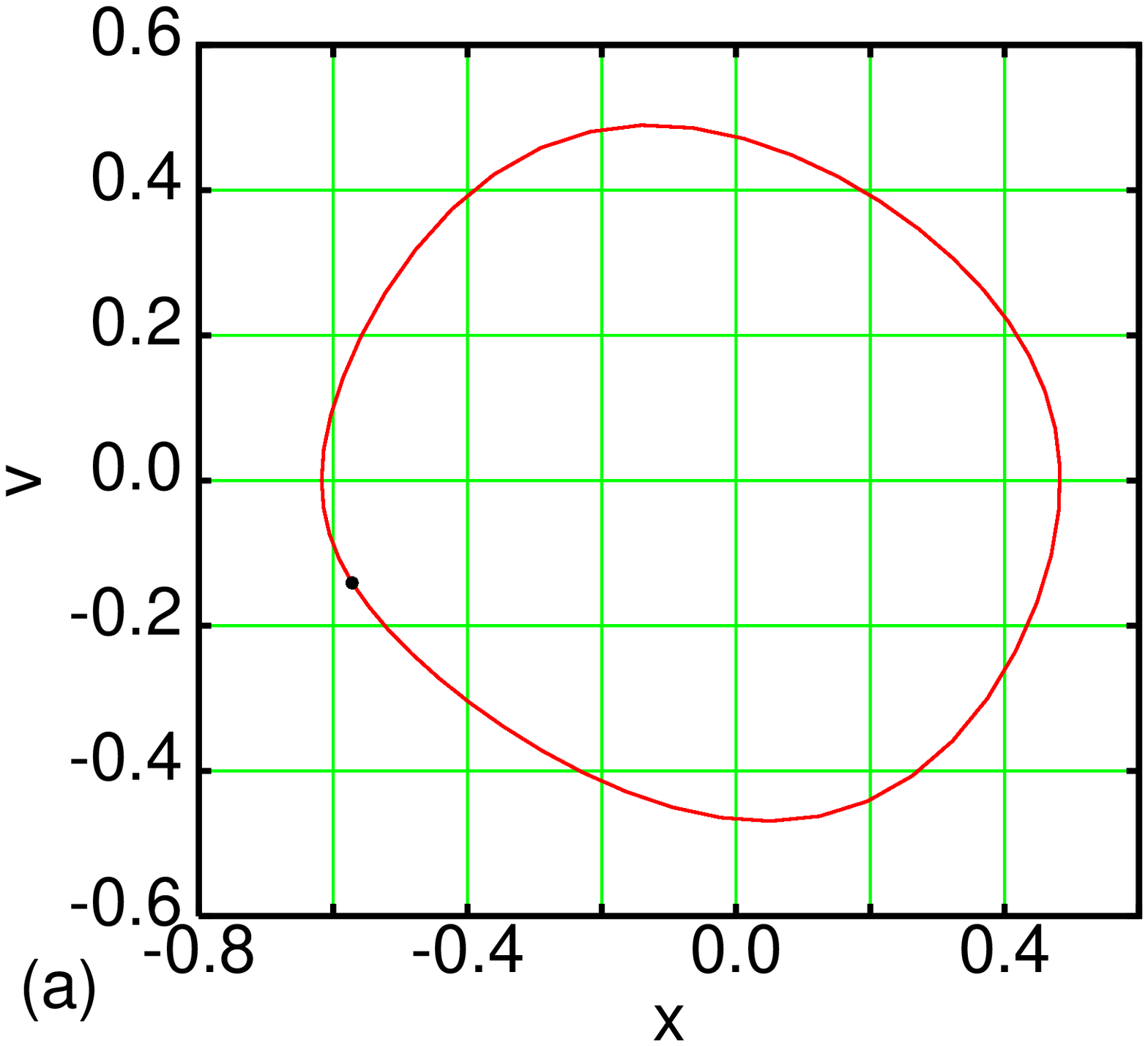,width=6.5cm,angle=-90}
\epsfig{file=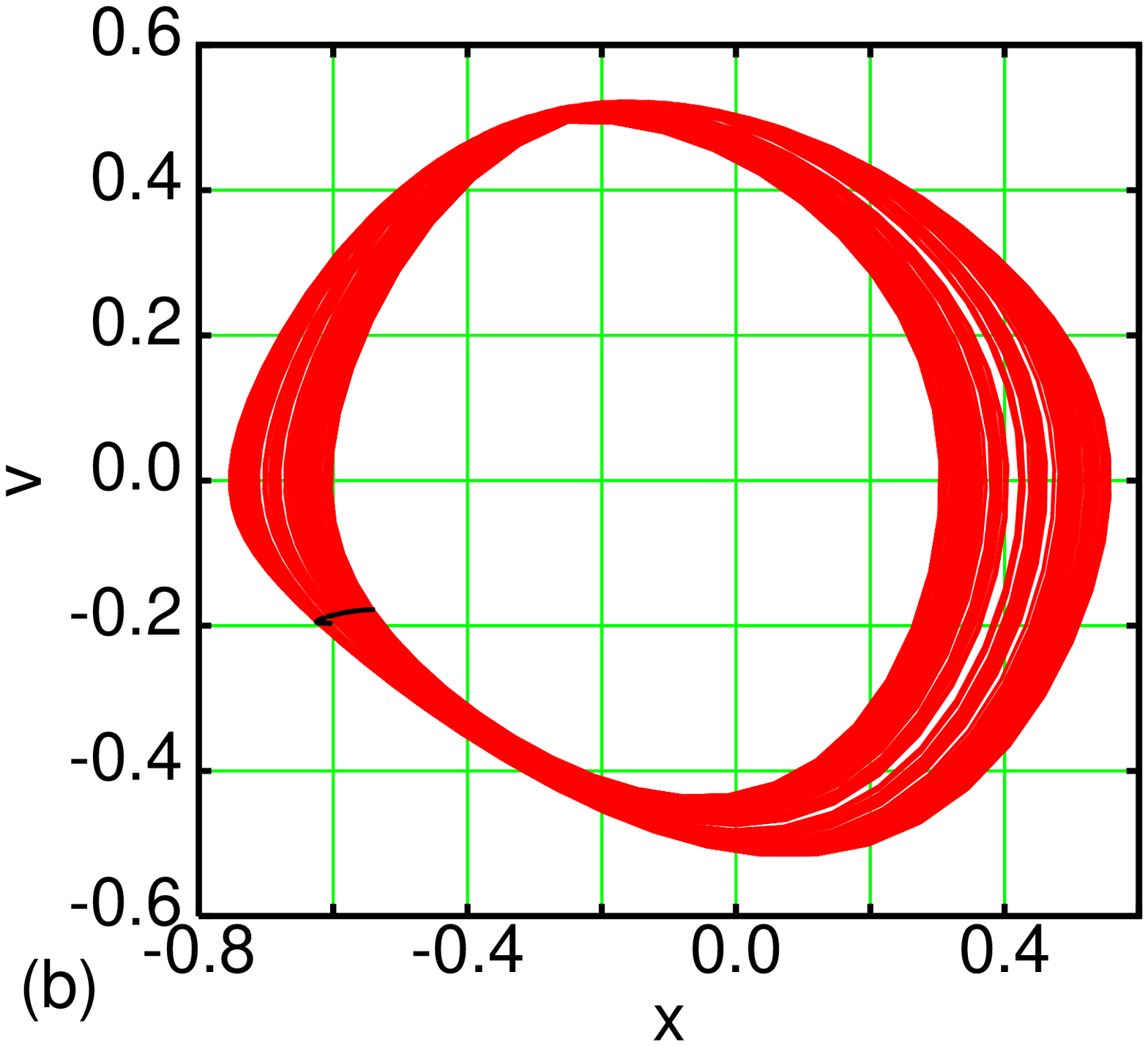,width=6.5cm,angle=-90}
}

\centerline{
\epsfig{file=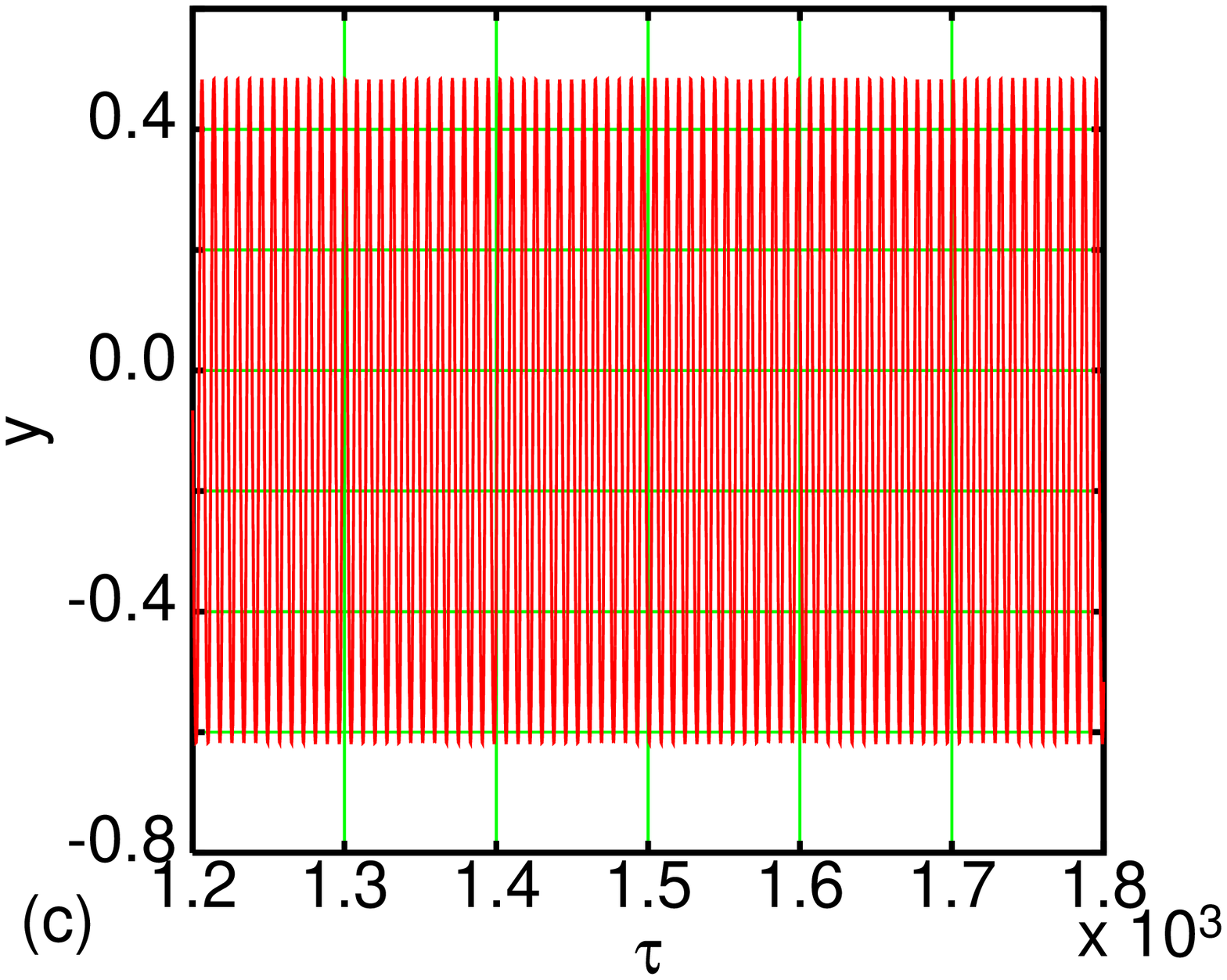,width=6.5cm,angle=-90}
\epsfig{file=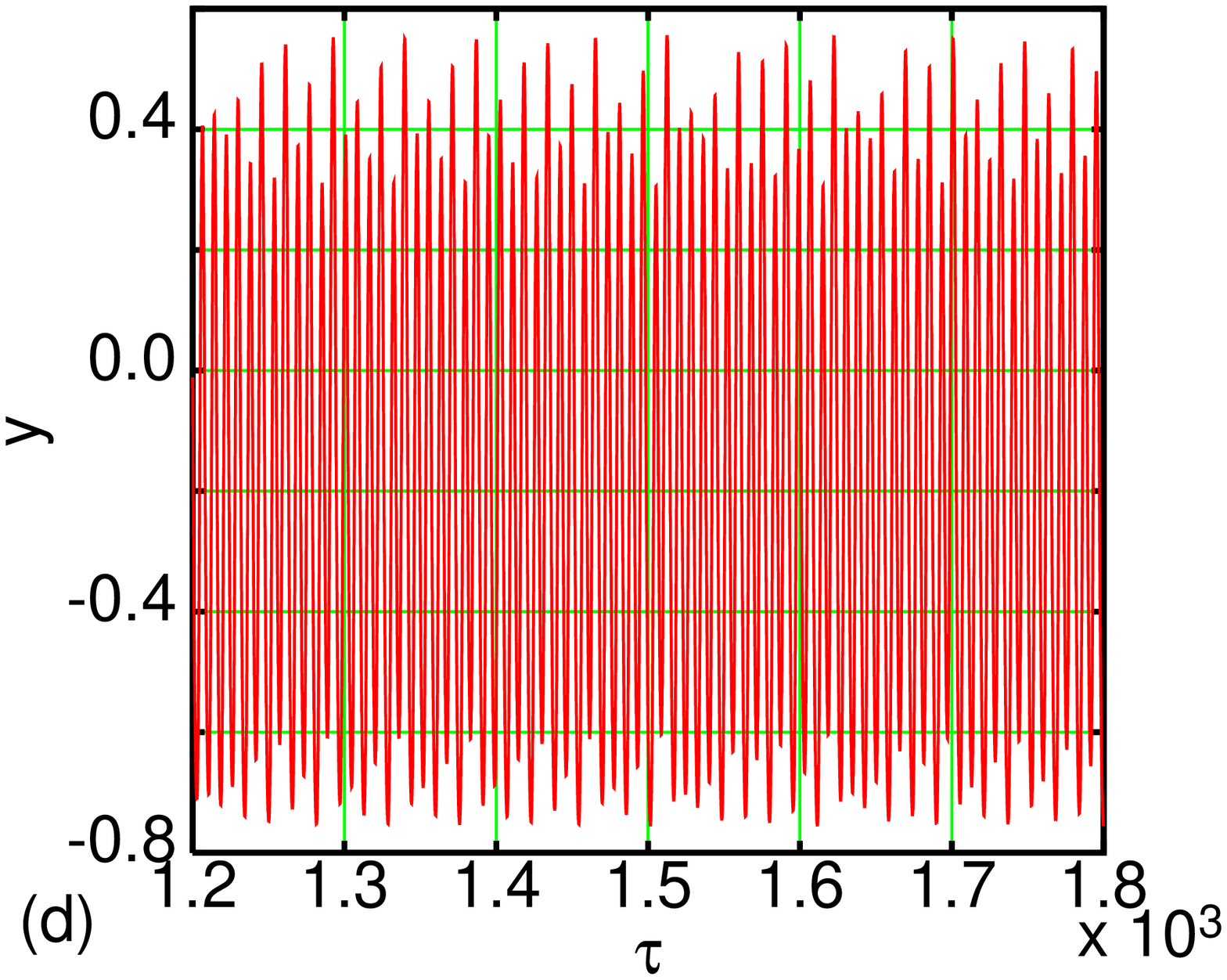,width=6.5cm,angle=-90}
}
\caption{ \label{fig9}
Phase portraits and Poincare maps below and above the critical amplitude at $\Omega'=0.8$
for regular (Fig. \ref{fig9}a, $A=0.31$m and initial conditions $[x_{in},v_{in}]$ = $[0.15,0.1]$)
and chaotic (Fig. \ref{fig9}b, $A=0.41$m and the initial conditions $[x_{in},v_{in}]$ = $[-0.15,0.1]$)
solutions.
The dominant Lyapunov exponents are $\lambda_1=-0.1625$ and
$\lambda_1=0.0354$, respectively.
Figs. \ref{fig9}c and \ref{fig9}d give the corresponding time responses.
Here $v \omega$ is given in m/s while $y$ is in m.
}
 \end{figure}

To illustrate the influence of a global homoclinic transition on the system dynamics,  
simulations were performed for interesting values of the system parameters,
using  Eq. (\ref{eq7}) for the model in Fig. \ref{fig1}. Knowing the critical value of the 
road profile amplitude $A_c$ (Fig. \ref{fig7}) and looking at a
 typical homoclinic bifurcation \cite{Tyrkiel2005}  
the effect on the resonance curves were examined first. Figure \ref{fig8} 
shows the sequence of resonance curves for $A=0.11$, 0.16, 0.21, 0.26, 0.31 and 0.36m 
respectively.
Apart from a typical shift of the maximum response to the right all of these curves 
are very similar, up to $A=0.31$m. For $A=0.41$m the synchronized 
solution is not stable in the region of resonance.
The other difference in the system behaviour occurs to the left side of the 
resonance peak where multiple solutions of the nonlinear system appear 
(in this case resonant and non-resonant solutions)
in the region of the resonance.
One can observe that starting from $A= 0.26$m the curves in Fig. \ref{fig8} show a
discontinuity signaling jumps 
between the resonant and non-resonant vibration amplitude $A_{OUT}$. Note in all cases 
a series of simulations were performed to calculate the system response, with $\Omega'$ 
decreasing as in Figs. \ref{fig3}a--b. For most of 
curves the same initial conditions were used for large $\Omega'$, namely $[x_{in},v_{in}]=[0.15,0.1]$.
However if the system escaped from the potential well initial conditions of 
$[x_{in},v_{in}]=[-0.15,0.1]$ and $[0,0.1]$ were used to avoid this effect.
For $A=0.36$m this was not possible in the vicinity of the resonance peak where 
the system escaped from the potential well for any initial conditions.
Moreover just before this escape (for $A=0.36$m $\approx A_c$) we observe a further increase 
in the vibration amplitude $A_{OUT}$.
Examining the related bifurcations diagrams 
we identified period doubling phenomenon occurring in this region which 
can be classified as a precursor of  
chaotic vibrations. Indeed alternative criteria to the Melnikov 
approach (Eq. \ref{eq24}) are based on the period doubling 
cascade \cite{Szemplinska1988,Kapitaniak1991}.
For larger amplitudes the unstable vibration region where escape from the potential is possible increases.
On the other hand, at $A=A_c$ the border between the basins of 
attraction belonging to different solutions disappears. To avoid these difficulties 
for further analysis the synchronized solution for $\Omega'=0.8$ at $A=0.31$m (Fig. \ref{fig8}) was used,
and then the excitation amplitude was increased slightly to $A=0.41$m, crossing the 
critical amplitude of $A_c \cong 0.35$m. Figures \ref{fig9}a-b show the  
phase portraits (by lines) and Poincare 
maps (by points) for these two cases. Figure \ref{fig9}a shows a synchronized motion 
while Fig. \ref{fig9}b corresponds to a chaotic attractor. The dominant Lyapunov 
exponents calculated for these responses 
are $\lambda_1=-0.1625$  (Fig. \ref{fig9}a) and $\lambda_1=0.03540$ (Fig. \ref{fig9}b).
Note, the chaotic attractor is very similar to that studied by Thompson \cite{Thompson1989} 
where the harmonic potential
has been supplemented by a nonlinear term with displacement to the power 3  ($z^3$). 
Figure \ref{fig9} also shows the time histories for the two cases:
$A=0.31$m (Fig. \ref{fig9}c) and $A=0.41$m (Fig. \ref{fig9}d). In this figure 
the difference between the periodic and chaotic responses is clear.

\section{Summary and Conclusions} 

We have studied the vibrations of a quarter-car model 
 with a softening stiffness of the Duffing type,
focusing on the potential for chaotic behaviour.
The model and parameters used were taken from the paper by Li {\em et al.} \cite{Li2004}, 
with the addition of the gravity force. The addition of this gravity force breaks the symmetry
of the potential, so that $V(-x) \neq V(x)$. 
The hysteretic nature of the damper caused a range of interesting system behaviour,
such as quasi-periodic, synchronized and chaotic motion. 
This had a substantial effect on the heteroclinic orbits, which transformed into homoclinic orbits.
We examined the global homoclinic bifurcations that appear as instabilities at the boundaries of  
the basins of attraction, and the cross-sections of stable and unstable manifolds,
by the perturbation approach using Melnikov theory.
A critical amplitude was found for which the system 
can exhibit chaotic vibrations. 
The analytic results have been confirmed by numerical simulations. 
In particular, the chaotic strange attractor was found for an excitation amplitude $A$
at the critical value, $A_c$, and a period doubling precursor for $A=0.36$m.
The transition to chaos appears to be present for $A_c \cong 0.36$m but could be lowered 
significantly for a smaller damping coefficient $c_2$ (Fig. \ref{eq7}).
Fortunately this region is beyond the usual amplitude of road profile excitation.
The chaotic solution appears just before the escape from the potential well, which is similar to
the system with a non-symmetric potential described by Thompson 
\cite{Thompson1989}.

\section*{Acknowledgements}
This research has been partially supported by the Polish Ministry of
Science and Information.

\newpage

\end{document}